\documentclass[runningheads]{llncs}
\usepackage[T1]{fontenc}
\usepackage{cite}
\usepackage{amsmath,amssymb,amsfonts}
\usepackage{graphicx}
\usepackage{textcomp}
\usepackage{xcolor}
\usepackage{graphicx}
\usepackage{bm}
\usepackage{multirow}
\usepackage{caption}
\usepackage{algorithm}
\usepackage{subfigure}
\usepackage{diagbox}
\usepackage{amssymb}
\usepackage{booktabs}
\usepackage{makecell}
\usepackage{adjustbox}
\usepackage{enumitem}
\usepackage{marvosym}
\usepackage{algorithmicx}
\usepackage{algpseudocode}
\usepackage{cite}
\usepackage{braket}
\usepackage{color}
\usepackage{ifsym}
\usepackage{graphicx}
\begin{document}
\title{An Enhanced Audio Feature Tailored for Anomalous Sound Detection Based on Pre-trained Models}
\titlerunning{An Enhanced Audio Feature Tailored for Anomalous Sound Detection}
% If the paper title is too long for the running head, you can set
% an abbreviated paper title here
%
\author{Guirui Zhong$^{1}$, Qing Wang$^{1\text{(\Letter)}}$, Jun Du$^{1}$, Lei Wang$^{2}$, \\ 
Mingqi Cai$^{3}$, and Xin Fang$^{3}$}
\authorrunning{G. Zhong et al.}
% First names are abbreviated in the running head.
% If there are more than two authors, 'et al.' is used.
%
\institute{
$^1$ University of Science and Technology of China, Hefei, China \\
\texttt{qingwang2@ustc.edu.cn} \\
$^2$ National Intelligent Voice Innovation Center, Hefei, China \\
$^3$ iFLYTEK Research, Hefei, China 
}
\maketitle              % typeset the header of the contribution
\begin{abstract}
Anomalous Sound Detection (ASD) aims at identifying anomalous sounds from machines and has gained extensive research interests from both academia and industry. However, the uncertainty of anomaly location and much redundant information such as noise in machine sounds hinder the improvement of ASD system performance. This paper proposes a novel audio feature of filter banks with evenly distributed intervals, ensuring equal attention to all frequency ranges in the audio, which enhances the detection of anomalies in machine sounds. Moreover, based on pre-trained models, this paper presents a parameter-free feature enhancement approach to remove redundant information in machine audio. It is believed that this parameter-free strategy facilitates the effective transfer of universal knowledge from pre-trained tasks to the ASD task during model fine-tuning. Evaluation results on the Detection and Classification of Acoustic Scenes and Events (DCASE) 2024 Challenge dataset demonstrate significant improvements in ASD performance with our proposed methods. 

\keywords{DCASE  \and Anomalous Sound Detection \and Feature Extraction \and Feature Enhancement.}
\end{abstract}

\section{Introduction}
\label{sec:intro}
The goal of Anomalous Sound Detection (ASD) is to detect whether the sound produced by a target machine is normal or anomalous. With the widespread use of machines, timely identification of mechanical faults is not only beneficial for machine maintenance, but also helps reduce property losses and protect personal safety. In factories, experienced workers can assess whether a machine is functioning normally or not based on the sound it emits. However, this method is highly subjective and may not consistently detect the machine's condition. Hence, developing an intelligent ASD system is essential to replace manual detection and improve reliability.

In recent years, the Detection and Classification of Acoustic Scenes and Events (DCASE) Challenge has attracted extensive attention from scholars in the fields of signal processing and machine learning, effectively driving advancements in acoustic scene classification, acoustic event detection, and related fields. The ASD task was first introduced into the competition in 2020 and has been held five times so far. The purpose of this task is to use machine learning models to detect whether a machine's sound is abnormal, thereby enabling intelligent monitoring of machine state\cite{nishida2024description,harada2023first}.

Due to the scarcity of abnormal sounds in realistic scenarios, only normal sounds emitted from machines can be used for model training, making ASD a self-supervised learning (SSL) task rather than a simple binary classification problem (i.e., normal or abnormal). To address this challenge, scholars have explored various auxiliary tasks to accomplish ASD. These auxiliary methods can be broadly divided into two main categories: discriminative learning and generative learning. Discriminative learning approach assumes that if models classify a specific machine type's sound as other types, this sound will be considered anomalous. Alternatively, generative learning approach evaluates anomalous likelihood based on reconstruction error: the higher the error, the more likely the sound is anomalous. Both methods operate on the premise that models trained solely on normal sounds struggle with anomalous audio due to distribution differences between normal and abnormal data.

Recently, methods based on pre-trained models have achieved significant progress compared to previous non-pre-trained approaches. These methods leverage a Vision Transformer (ViT) backbone pre-trained on AudioSet and fine-tune it on machine audio, aiming to transfer the model's general audio processing capabilities to the specific domain of machine audio\cite{jiang2024anopatch,LvAITHU2024}. However, there are still some obstacles that impede further performance improvements in ASD. Firstly, due to diverse causes of machine anomalies, it is difficult to determine the exact position of abnormal sounds on the spectrogram, which means that anomalous sounds can appear across different frequency ranges of the spectrogram. Furthermore, machine sounds often contain a great deal of trivial information, making it challenging for models to effectively learn machine-related features.

To address the first challenge, we use Mel filter banks with evenly distributed intervals to ensure equal attention across all frequency ranges in machine sounds. Since machine anomalies can arise from various causes, they may appear in different frequency ranges. For the second challenge, a straightforward solution is to incorporate attention modules into pre-trained models. Nevertheless, training new module parameters from scratch in a pre-trained model often yields minimal benefits or can even be detrimental due to knowledge inconsistency between the newly added parameters and the existing pre-trained ones. To mitigate this issue and enhance ASD performance, we instead employ a parameter-free attention module. We highlight three key contributions of this paper as follows:
\begin{itemize}
\item[-] We design a novel audio feature to ensure equal attention across all frequency ranges and better capture the anomaly in machine sounds. \par
\item[-] We propose a parameter-free feature enhancement framework to remove redundant information and strengthen the relevant feature in machine audio. \par
\item[-] Evaluation results on the DCASE 2024 ASD dataset demonstrate significant improvements in ASD performance with our proposed methods. As a result, our system reaches a new state-of-the-art (SOTA) performance on the DCASE 2024 ASD evaluation dataset.\par
\end{itemize}

\section{Related Works}
\label{sec:related-work}
\subsection{Discriminative Learning}
Although anomalous sound detection is trained exclusively on normal machine sounds, these sounds contain categorical information such as machine type, domain, and attribute. The discriminative learning method leverages this information to accurately classify normal sounds, enabling the model to effectively characterize normal audio features and learn meaningful representations through classification tasks (e.g., machine type or machine attributes). During training, the model continuously optimizes its classification accuracy for normal data. As a result, during testing, sounds that cannot be correctly classified into the expected machine type or attribute are more likely to be identified as anomalous.

Wilkinghoff\cite{wilkinghoff2024self} explores a simple yet effective SSL approach for ASD, based on a convolutional neural network (CNN) that integrates both time-domain and frequency-domain information from audio. Wang et al.\cite{wang2024representation} introduce a two-stage multi-attribute classification framework tailored for ASD in real scenarios, facilitating the extraction of discriminative representations of machine sounds. Recently, Jiang et al.\cite{jiang2024anopatch} propose to utilize a ViT backbone pre-trained on AudioSet and fine-tune it on machine audio, which has made great progress in ASD. Notably, the first-place entries in the DCASE ASD challenge have predominantly relied on discriminative learning method, highlighting its effectiveness in ASD.

\subsection{Generative Learning}
The generative learning method is one of the earliest approaches used to detect anomalous sounds, with auto-encoder (AE) being a typical example. AE reconstructs the original signal by encoding and decoding the input sound signal through a neural network. Anomalous sounds are then detected based on the reconstruction error. Specifically, for normal sounds, because we have seen the normal sound samples when training the AE, the signal recovered by encoding and decoding closely resembles the input signal. In contrast, for anomalous sounds, which the AE has not encountered during training and which differ from normal sounds, the reconstructed signal deviates significantly from the original input. This discrepancy enables the detection of anomalous sounds.

Zhou et al.\cite{zhou2023autoencoder} design an auxiliary classifier into AE in a multi-task learning manner and a group-based decoder structure to enhance ASD. Guan et al.\cite{guan2023transformer} introduce an ID-constrained Transformer-based AE architecture, which mitigates the generalization of AE for anomalous sounds and enhances the distinguishing ability for different machines of the same type. Moreover, Jiang et al.\cite{jiang2023unsupervised} utilize a generative adversarial network (GAN) to detect anomalous sounds.

\section{Proposed Method}

\begin{figure}[t]
	\centering
	\includegraphics[width=1.0\linewidth]{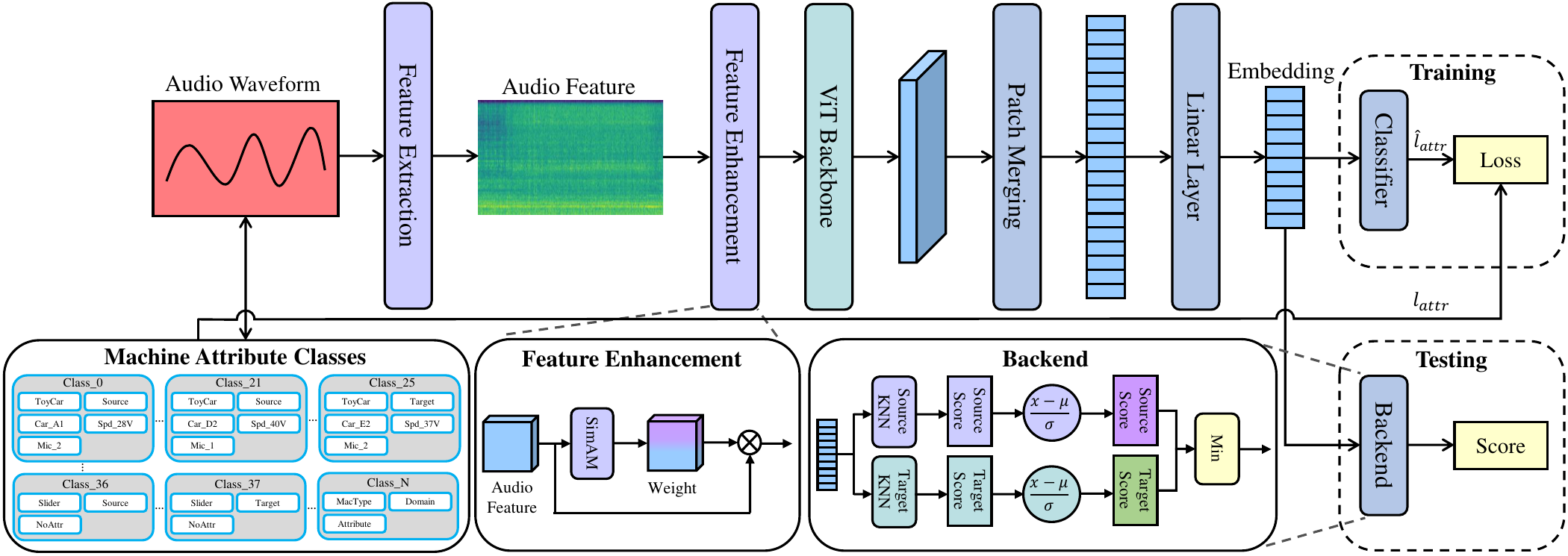}
    \vspace{-0.5cm}
	\caption{The proposed ASD model consists of a modified feature extraction module, a feature enhancement module, a network backbone to classify machine attributes, and a backend with domain normalization technique.}
    \label{fig: simam}
    \vspace{-0.5cm}
\end{figure}

The overall flowchart of the proposed ASD framework is illustrated in Fig. \ref{fig: simam}, mainly consisting of a modified feature extraction module, a feature enhancement module, a ViT backbone, and a backend with domain normalization technique. The feature extraction module uses a set of evenly distributed filter banks to extract the audio feature based on the conventional process of extracting FBank. Meanwhile, as shown in the lower middle box of Fig. \ref{fig: simam}, the feature enhancement module comprises an attention module to obtain weights using the Simple, Parameter-Free Attention Module (SimAM)\cite{yang2021simam}. Then, in the bottom-right box, we introduce K-Nearest Neighbors (KNN) detectors with a simple domain normalization strategy to generate the anomaly score, which is less affected by domain shift. By means of these techniques and the network backbone, we carry out machine attribute classification task to realize ASD. These proposed methods are elaborated in the following subsections.
\subsection{Feature Extraction}
In this section, we review the extraction process of the original filter banks (FBank) feature. The process can be mainly divided into the following steps: pre-emphasis, framing, applying window functions, Fast Fourier Transform (FFT), Mel filtering, and logarithmic operation. As frequency increases, the spacing between adjacent Mel filters gradually widens when Mel filtering. This design prioritizes low-frequency information, aligning with the human ear’s greater sensitivity to low frequencies and reduced sensitivity to high frequencies.

However, this Mel filter intervals arrangement does not apply to the ASD task. On the one hand, some machine sounds are primarily distributed in higher frequency ranges. On the other hand, due to the unpredictability of real-world scenarios, there are many different reasons that can make the machine anomalous, causing that the anomaly will occur at random, unfixed frequency ranges. To address this, we propose a modified FBank feature tailored for ASD task by adjusting the intervals between adjacent Mel filters to be evenly spaced. In other words, all adjacent Mel filters are distributed uniformly to enhance anomaly detection in machine sounds. Notably, except for the change of intervals, the process of extracting modified FBank features is consistent with the original process.

\subsection{Feature Enhancement}
\subsubsection{Parameter-free Attention Module}
Due to the presence of redundant information in sounds, such as background noises, it is necessary to enhance the features of machine sounds to enable the model to focus on crucial information. One straightforward and widely used approach is to incorporate attention modules within the network. These attention modules can be categorized into two classes: the first class involves designing specific trainable layers distinct from the backbone network, known as parameter-requiring modules; the second class involves obtaining the corresponding weights through elaborate formulas, which usually do not require training, referred to as parameter-free modules.

Because we use pre-trained models, which have been proven to be more effective in contrast with previous non-pre-trained ways, as our baseline, parameter-free attention modules are applied in the network as the feature enhancement layer. This choice serves two key purposes. On the one hand, it is usually unreasonable to add new trainable parameters to pre-trained models, as there is knowledge inconsistency between the newly added parameters and the existing pre-trained parameters, making it difficult for the attention modules to perform their expected functions. Notably, this differs from the common practice of replacing the pre-trained decoder with specific modules for downstream tasks during fine-tuning. For instance, linear layers are used to replace the original decoder to perform downstream classification tasks. However, attention modules are generally placed within or before encoder, with only slight adjustments made during fine-tuning. On the other hand, even though it is feasible to freeze the pre-trained parameters and then fine-tune the newly added attention parameters to achieve knowledge consistency, this method is still more complex than directly adding parameter-free attention modules and fine-tuning the model only once, and it cannot guarantee effectiveness.
\subsubsection{Feature Enhancement Module}
For the feature enhancement layer, we deploy SimAM as the attention module. SimAM assigns a corresponding weight to each neuron, calculated using a formula based on the spatial suppression phenomenon. The SimAM calculates the respective weights of all neurons on a single channel. Thus, giving an audio spectrogram $\mathbf{X}\in\mathbb{R}^{1\times{F}\times{T}}$, where $F$ denotes the frequency dimension and $T$ denotes the time dimension, we refer to the weights obtained from the SimAM as $\mathbf{W}\in\mathbb{R}^{1\times{F}\times{T}}$. Given the target neuron $z$ and other neurons $x_{i}$ in $\mathbf{X}$, the process of SimAM can be defined below:
\begin{gather}
\label{eq: simam}
    {e}_{z}^{\ast}= \frac{4(\hat{\sigma}^{2}+\lambda)}
    {(z-\hat{\mu})^{2}+2\hat{\sigma}^{2}+2\lambda} \\
    {w}_{z}^{\ast}= \mathrm{sigmoid}({1}/{{e}_{z}^{\ast}})
\end{gather}
where $\hat{\mu}=\frac{1}{M}\sum_{i=1}^{M}x_{i}$, $\hat{\sigma}^{2}=\frac{1}{M}\sum_{i=1}^{M}(x_{i}-\hat{\mu})^{2}$, $\lambda$ is a hyperparameter, and $M=F\times T$ is the number of neurons on the single channel of $\mathbf{X}$. It is noted that we set $\lambda=10^{-4}$ (the default value of the original paper) in all experiments and do not tune it.  ${e}_{z}^{\ast}$ and ${w}_{z}^{\ast}$ represent the minimal energy and the weight of target neuron $z$; the lower energy ${e}_{z}^{\ast}$, the greater the weight ${w}_{z}^{\ast}$. $\mathrm{sigmoid}(\cdot)$ is the sigmoid activation function to restrict too large value. Finally, we obtain $\mathbf{W}$ by grouping all target neurons weights ${w}_{z}^{\ast}$ across the channel. In fact, the weight represents the linear separability between the target neuron and other neurons; the higher the linear separability, the greater the weight. Specific derivation process can be found in \cite{yang2021simam}.

\subsection{KNN Backend with Domain Normalization}
\label{ssec: DN}
In real-world scenarios, machine operational states and the environmental noises are not constant, which causes the domain shift problem and greatly restricts the performance of the ASD system. To address this, we introduce a simple domain normalization strategy in the backend KNN detector to alleviate the problem during inference, building upon a similar method from \cite{ZhaoCUMT2024}. The concrete operation is presented in the bottom-right box of Fig. \ref{fig: simam}.

To start with, we train two KNN detectors for each machine type, one using the embeddings from all samples in the source domain and the other using the embeddings from all samples in the target domain. Subsequently, two anomaly scores for a given testing machine audio are obtained by computing the cosine distance between the embedding of the testing audio and its closest neighbor (k = 1). Then, score normalization is performed between the two domains. In the end, the minimum anomaly score for each testing machine audio is adopted as the audio's final score. The aforementioned process is shown below:
\begin{gather}
\label{eq: score_st}
    \mathrm{Score}_{i}= \mathrm{d}({\mathbf{E}}_{{test}}, {\mathbf{E}}_{{i}}) \\
\label{eq: sn}
    \mathrm{Score}_{i}^{\mathrm N}= (\mathrm{Score}_{i} - \mu_{i}) / \sigma_{\mathrm i} \\
\label{eq: min}
    \mathrm{Score} = \mathrm{Min}(\mathrm{Score}_{s}^{\mathrm N}, \mathrm{Score}_{t}^{\mathrm N})
\end{gather}
where $\mathrm{d}(\cdot)$ denotes the cosine distance, the subscript `$i$' is the domain indicator, $\mathrm{Score}_{i}$ represents the anomaly score of the testing audio in the source (${i}={s}$) or target (${i}={t}$) domain, and ${\mathbf{E}}_{{test}}$ denotes the test embedding of machine audio acquired by the fine-tuned model. $\mathrm{Score}_{i}^{\mathrm N}$ represents normalized anomaly score. Variables $\mu_{i}$ and $\sigma_{i}$ indicate the mean and standard deviation of anomaly scores in the source or target domain. $\mathrm{Min}(\cdot)$ is the minimum operation.

\subsection{Machine Attribute Classification}
\label{ssec: Model}
In this section, we explain the whole process of ASD system used in this paper. The machine attribute classification schematic diagram can be found in Fig. \ref{fig: simam}. To begin with, the machine audio waveform $\mathbf{x}$ is transformed into the modified FBank feature $\mathbf{X}$ by the feature extraction module. What this feature extraction process distinguishes from the original FBank extraction process is the use of evenly distributed filter banks. Subsequently, this audio feature is input to the feature enhancement layer to remove redundant information in the feature. So far, we have obtained the enhanced audio feature $\mathbf{X}_{e}$ tailored for ASD system, the process can be defined below:
\begin{gather}
\label{eq: feature}
    \mathbf{X}= \mathrm{Extract}(\mathbf{x}) \\
\label{eq: enhance}
    \mathbf{X}_{e}= \mathrm{SimAM}(\mathbf{X}) \times \mathbf{X}
\end{gather}
where $\mathbf{X}$ can be original or modified FBank feature as described above, $\mathrm{Extract}(\cdot)$ represents the feature extraction process, and $\mathrm{SimAM}(\cdot)$ denotes the used attention module in the feature enhancement layer.

Then, the enhanced audio feature is split into multiple patches and input into the ViT backbone pretrained on the AudioSet\cite{gemmeke2017audio}. The backbone models each patch and outputs embeddings for all patches. Thereafter, the attentive statistics pooling layer\cite{okabe2018attentive} is used to merge all patches information into an utterance embedding. Lastly, the utterance embedding is mapped into low-dimensional embedding. This embedding is utilized for machine attribute classification by ArcFace\cite{deng2019arcface} classifier during fine-tuning. We combine each machine type, domain, and its corresponding attributes as a separate class and perform classification tasks. For those machine types without attributes, we combine their machine types and domains as separate classes. The details can be found in the dataset introduction section later. After fine-tuning, this embedding is used for anomaly detection in the KNN backend with the domain normalization technique during testing. The training stage can be defined below:
\begin{gather}
\label{eq: embedding}
    \mathbf{E} = \mathcal{F}(\mathbf{X}_{e}) \\
\label{eq: arcface}
    \hat{l}_{attr} = \mathcal{C}_{attr}(\mathbf{E}) \\
\label{eq: loss}
    \mathcal{L}_\mathrm{ASD} = \mathrm{CE}(\hat{l}_{attr}, {l}_{attr})
\end{gather}
where $\mathbf{E}$ and $\mathcal{F}(\cdot)$ denote the embedding of machine audio and the map function including ViT, patch merging, and linear mapping operation, separately. $\mathcal{C}_{attr}(\cdot)$ indicates the ArcFace attribute classifier, $\mathrm{CE}(\cdot)$ is the cross-entropy loss, and $\mathcal{L}_\mathrm{ASD}$ is the ASD loss to be optimized. ${l}_{attr}$ and $\hat{l}_{attr}$ represent the ground truth and predicted attribute labels. After training, the embedding will pass through backend to obtain anomaly score.
\section{Experiments}

\subsection{Dataset and Metrics}
\begin{figure}[t]
	\centering
	\includegraphics[width=0.9\linewidth]{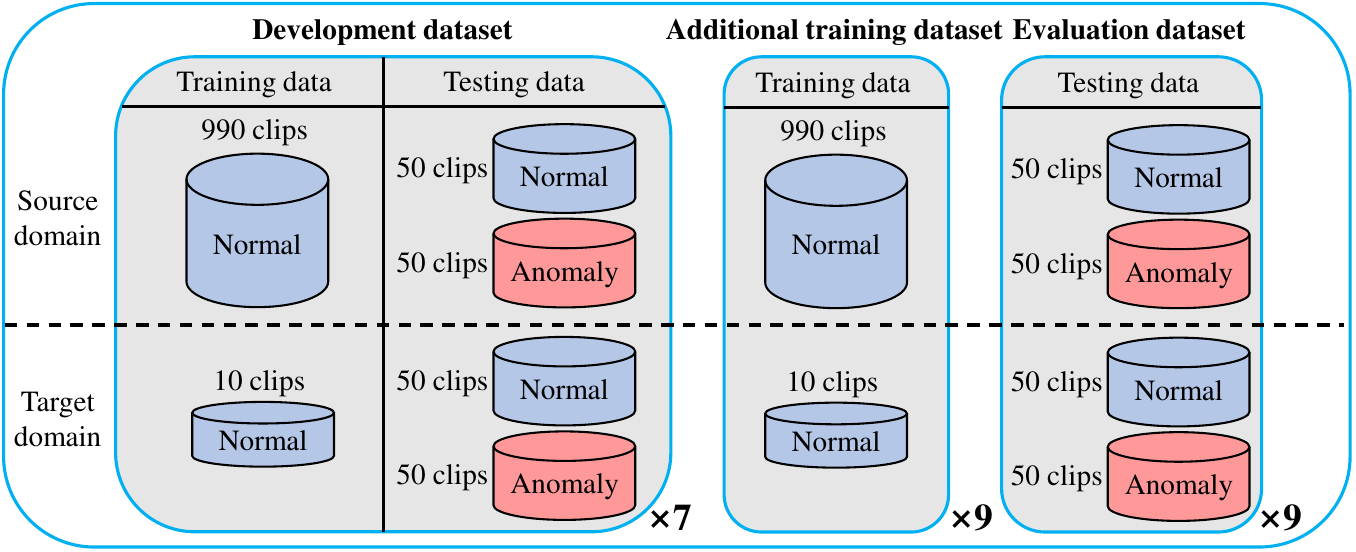}
    \vspace{-0.2cm}
	\caption{The structure of the DCASE 2024 ASD dataset.}
    \label{fig: dataset}
    \vspace{-0.5cm}
\end{figure}

We evaluate the ASD task on the DCASE 2024 ASD dataset\cite{harada2021toyadmos2,dohi2022mimii}, a widely accepted public benchmark for many papers in the ASD area. The structure of the dataset can be found in Fig. \ref{fig: dataset}. It is noted that there are seven and nine machine types in the development and additional (including additional training and evaluation) datasets, respectively. Each machine type has 1000 clips for training and 200 clips for testing. Some machine types have corresponding attribute information as metadata. For instance, as shown in the bottom-left box of Fig. \ref{fig: simam}, the ToyCar has car model (Car), speed (Spd), and microphone number (Mic) attributes. Unique value of attributes, combining with machine type and domain, is considered as a separate class. Additionally, for those machine types without attributes, we combine their machine types and domains as separate classes. We evaluate all methods with the area under the receiver operating characteristic (ROC) curve (AUC) and the partial AUC (pAUC) using the anomaly score that indicates the anomaly degree. Meanwhile, we compare the harmonic mean performance on the development set, evaluation set and all machine types, which we all refer to as score. The harmonic mean calculated on the evaluation set serves as the official score according to the challenge rules\cite{nishida2024description,harada2023first}.

\subsection{Implementation Details}
The overall framework used in this work is based on the public open-source Efficient Audio Transformer (EAT) project\cite{chen2024eat}. All audio waveforms are padded or truncated to 10s, and then converted to log-mel spectrograms with a frame length of 25ms, a frame shift of 10ms, and 128 mel bins. We use the Adam optimizer, and the pre-trained EAT is fine-tuned for appropriate epochs to converge with a batch size of 32. A cosine learning rate scheduler is adopted with an upper limit of 5e-5 and a warm-up step of 120. Mixup\cite{zhang2017mixup} and SpecAugment\cite{park2019specaugment} data augmentation techniques are utilized to generate more diverse data during the training step with mixup $\alpha$ of 0.8 and mask ratio of 0.2 on the time and frequency axes, respectively. We execute all experiments 5 times independently and report the mean and standard deviation.

\subsection{Experimental Results and Analysis}
\subsubsection{Feature Extraction Module}
We use the EAT pre-trained on the AudioSet with original FBank feature and fine-tune it using different FBank features. In the fine-tuning stage, we validate the effect of different audio features by modifying the distribution of filter banks. We refer to original FBank and modified FBank as o-FB and m-FB, separately. Meanwhile, we also compare another filter bank, Gammatone filter, referred to as g-FB. AUC$_{\mathrm s}$, AUC$_{\mathrm t}$, and pAUC represent the AUC in the source domain, the AUC in the target domain, and the pAUC across both domains. From Table \ref{tab: Extraction}, we can find that using m-FB to fine-tune the model achieves the best performance compared with others, which aligns with our motivation for designing the evenly distributed filter banks, i.e., better capturing the anomalies in machine sounds. Besides, the reason for g-FB's performance is that it is too different from o-FB used in pre-training, which makes the model unable to adapt in fine-tuning, while m-FB has good adaptability.

\subsubsection{Domain Normalization Method}
As shown in Table \ref{tab: Extraction}, it can be found that greater performance is achieved by using domain normalization. Upon closer inspection of the AUC values between the two domains, it is evident that the domain normalization technique primarily enhances the performance by bridging and narrowing the gap between them. This aligns with our motivation for designing the domain normalization method, which aims to alleviate the domain shift problem. Due to the higher score, we use the m-FB feature and domain normalization method for later experiment validation.

\begin{table}[t]
	\renewcommand\arraystretch{1.25}
	\newcolumntype{L}[1]{>{\raggedright\arraybackslash}p{#1}}
	\newcolumntype{C}[1]{>{\centering\arraybackslash}p{#1}}
	\newcolumntype{R}[1]{>{\raggedleft\arraybackslash}p{#1}}
	\centering
	\caption{Performance comparison for different audio features and domain normalization (DN) method. Mean and standard deviation over five runs are reported.}
	\vspace{-0.1cm}
	\label{tab: Extraction}\medskip
	\resizebox{12 cm}{!}{\begin{tabular}{c c c c c c|c c c c|c}
			\toprule[1 pt]
			\multirow{2}{*}{Feature} & \multirow{2}{*}{DN} & \multicolumn{4}{c}{Development set} & \multicolumn{4}{c}{Evaluation set} & All \\
            {} & {} & AUC$_{\mathrm s}$ & AUC$_{\mathrm t}$ & pAUC & score & AUC$_{\mathrm s}$ & AUC$_{\mathrm t}$ & pAUC & score & score \\
            \midrule
            \multirow{2}{*}{o-FB} & - 
            & 78.21{\scriptsize$\pm$1.04} 
            & 56.85{\scriptsize$\pm$0.35} 
            & 56.47{\scriptsize$\pm$0.83} 
            & 62.39{\scriptsize$\pm$0.37} 
            & 71.43{\scriptsize$\pm$2.06} 
            & 62.11{\scriptsize$\pm$0.33} 
            & 57.91{\scriptsize$\pm$0.48} 
            & 63.33{\scriptsize$\pm$0.82} 
            & 62.91{\scriptsize$\pm$0.29}  \\
            {} & $\checkmark$ 
            & 69.64{\scriptsize$\pm$1.19} 
            & 67.69{\scriptsize$\pm$0.44} 
            & 55.65{\scriptsize$\pm$1.07} 
            & 63.68{\scriptsize$\pm$0.56} 
            & 66.13{\scriptsize$\pm$1.05} 
            & 72.16{\scriptsize$\pm$0.81} 
            & 57.02{\scriptsize$\pm$0.34} 
            & 64.49{\scriptsize$\pm$0.59} 
            & 64.13{\scriptsize$\pm$0.12}  \\
            \midrule
            \multirow{2}{*}{g-FB} & - 
            & 70.10{\scriptsize$\pm$0.37} 
            & 49.85{\scriptsize$\pm$0.89} 
            & 55.36{\scriptsize$\pm$0.55} 
            & 57.26{\scriptsize$\pm$0.45} 
            & 70.79{\scriptsize$\pm$1.13} 
            & 48.05{\scriptsize$\pm$1.40} 
            & 53.43{\scriptsize$\pm$0.46} 
            & 55.90{\scriptsize$\pm$0.74} 
            & 56.48{\scriptsize$\pm$0.22}  \\
            {} & $\checkmark$ 
            & 64.57{\scriptsize$\pm$0.33} 
            & 63.78{\scriptsize$\pm$1.04} 
            & \textbf{56.70}{\scriptsize$\pm$0.53} 
            & 61.47{\scriptsize$\pm$0.44} 
            & 62.27{\scriptsize$\pm$1.02} 
            & 63.18{\scriptsize$\pm$0.37} 
            & 52.71{\scriptsize$\pm$0.09} 
            & 58.98{\scriptsize$\pm$0.33} 
            & 60.04{\scriptsize$\pm$0.09}  \\
            \midrule
            \multirow{2}{*}{m-FB} & -
            & \textbf{78.38}{\scriptsize$\pm$0.74} 
            & 59.01{\scriptsize$\pm$0.40} 
            & 55.08{\scriptsize$\pm$0.58} 
            & 62.68{\scriptsize$\pm$0.14} 
            & \textbf{74.44}{\scriptsize$\pm$1.73} 
            & 64.24{\scriptsize$\pm$0.55} 
            & \textbf{59.03}{\scriptsize$\pm$0.41} 
            & 65.29{\scriptsize$\pm$0.55} 
            & 64.12{\scriptsize$\pm$0.25}  \\
            {} & $\checkmark$ 
            & 70.76{\scriptsize$\pm$0.43} 
            & \textbf{68.93}{\scriptsize$\pm$0.48} 
            & 54.73{\scriptsize$\pm$0.55} 
            & \textbf{63.95}{\scriptsize$\pm$0.28} 
            & 70.01{\scriptsize$\pm$0.11} 
            & \textbf{73.78}{\scriptsize$\pm$0.42} 
            & 58.05{\scriptsize$\pm$0.52} 
            & \textbf{66.57}{\scriptsize$\pm$0.32} 
            & \textbf{65.40}{\scriptsize$\pm$0.11}  \\     
			\bottomrule[1 pt]
	\end{tabular}}
	\vspace{-0.5 cm}
\end{table}

\begin{figure}[h]
	\centering
	\includegraphics[width=1.0\linewidth]{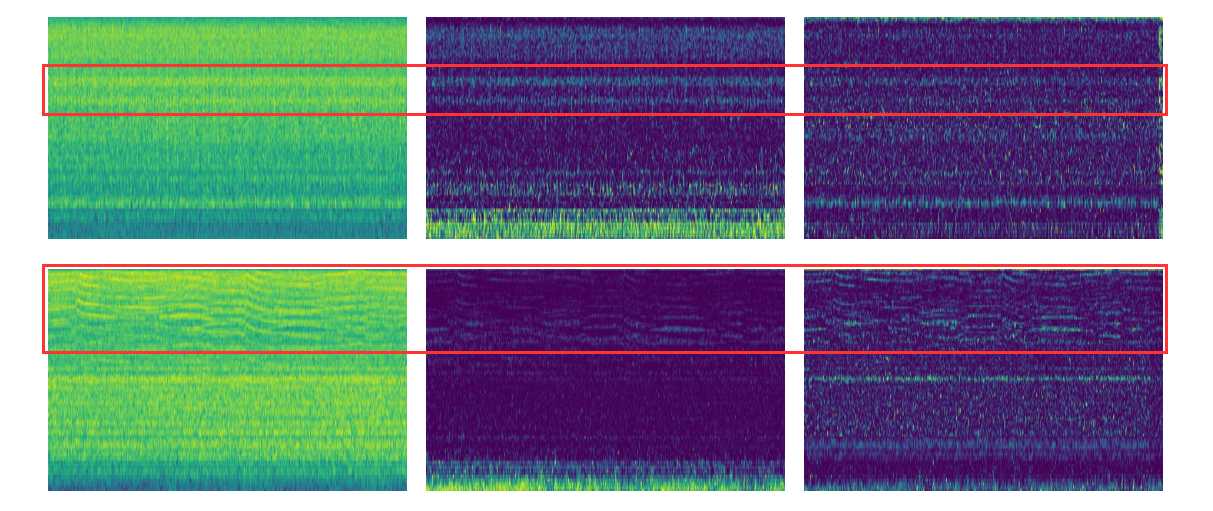}
    \vspace{-0.5cm}
	\caption{Visualization of the audio spectrogram, the global weight from the original SimAM, and the local weight obtained by the improved SimAM, are shown in the left, middle, and right sides, respectively. The first and second rows represent BrushlessMotor and HoveringDrone machines, respectively. The key distinguishing features of different machine types are framed in the red rectangular boxes.}
    \label{fig: simam2}
    \vspace{-0.5cm}
\end{figure}

\subsubsection{Feature Enhancement Module}
We validate the effect of feature enhancement module in the way of the last row of Table \ref{tab: Extraction}. We first apply the original SimAM, which is a form of global feature enhancement from the view of the whole spectrogram, to the audio spectrogram, which only leads to improvements of 0.28\% on all machine types. After further analysis of this phenomenon, as shown in the Fig. \ref{fig: simam2}, we know that from the global view, the characteristic lines of some machine types cannot be well captured because their characteristic lines are small or complex and easily disturbed by noise. Therefore, we split the whole spectrogram into multiple small spectrograms with a size of ${32\times32}$ to utilize feature enhancement from the local view. As shown in the third row of Table \ref{tab: Enhancement}, the local feature enhancement has further improvements on the overall performance (all score). From spectrograms of different machine types and experimental results, we can find that some machine types and others are suitable from the local and global views, respectively. Moreover, machine condition detection is performed under a specific machine type, which is known to us during both training and testing. This distinguishes ASD approaches from other classification tasks. Thus, we further utilize the prior knowledge of machine types to apply a certain feature enhancement method (global or local) for each machine type, which we call customized feature enhancement. Specifically, as shown in the Fig. \ref{fig: simam2}, global feature enhancement is applied to machine types such as BrushlessMotor, which have simple and easily identifiable feature lines, while local feature enhancement is performed on types like HoveingDrone, which exhibit fine and difficult-to-detect feature lines. As shown in the last row of Table \ref{tab: Enhancement}, the customized feature enhancement approach gains further improvements and achieves the best performance among all feature enhancement modes.

\begin{table}[t]
	\renewcommand\arraystretch{1.25}
	\newcolumntype{L}[1]{>{\raggedright\arraybackslash}p{#1}}
	\newcolumntype{C}[1]{>{\centering\arraybackslash}p{#1}}
	\newcolumntype{R}[1]{>{\raggedleft\arraybackslash}p{#1}}
	\centering
	\caption{Performance comparison for different feature enhancement modes. Mean and standard deviation over five runs are reported.}
	\vspace{-0.1cm}
	\label{tab: Enhancement}\medskip
	\resizebox{12 cm}{!}{\begin{tabular}{c c c c c|c c c c|c}
			\toprule[1 pt]
			\multirow{2}{*}{Mode} & \multicolumn{4}{c}{Development set} & \multicolumn{4}{c}{Evaluation set} & All \\
            {} & AUC$_{\mathrm s}$ & AUC$_{\mathrm t}$ & pAUC & score & AUC$_{\mathrm s}$ & AUC$_{\mathrm t}$ & pAUC & score & score \\
            \midrule
            Vanilla & 70.76{\scriptsize$\pm$0.43} 
            & 68.93{\scriptsize$\pm$0.48} 
            & 54.73{\scriptsize$\pm$0.55} 
            & 63.95{\scriptsize$\pm$0.28} 
            & 70.01{\scriptsize$\pm$0.11} 
            & 73.78{\scriptsize$\pm$0.42} 
            & 58.05{\scriptsize$\pm$0.52} 
            & 66.57{\scriptsize$\pm$0.32} 
            & 65.40{\scriptsize$\pm$0.11}  \\
            \midrule
            Global & 72.02{\scriptsize$\pm$1.25} 
            & 69.11{\scriptsize$\pm$1.07} 
            & 55.75{\scriptsize$\pm$0.17} 
            & 64.80{\scriptsize$\pm$0.51} 
            & \textbf{70.57}{\scriptsize$\pm$0.80} 
            & 72.65{\scriptsize$\pm$1.00} 
            & 57.95{\scriptsize$\pm$0.39} 
            & 66.38{\scriptsize$\pm$0.24} 
            & 65.68{\scriptsize$\pm$0.15}  \\
            \midrule
            Local & \textbf{72.87}{\scriptsize$\pm$1.51} 
            & 70.32{\scriptsize$\pm$0.52} 
            & 56.86{\scriptsize$\pm$0.99} 
            & 65.88{\scriptsize$\pm$0.78} 
            & 69.77{\scriptsize$\pm$1.31} 
            & 73.00{\scriptsize$\pm$0.55} 
            & 57.77{\scriptsize$\pm$0.26} 
            & 66.16{\scriptsize$\pm$0.57} 
            & 66.03{\scriptsize$\pm$0.39}  \\
            \midrule
            Customized & 72.25{\scriptsize$\pm$0.89} 
            & \textbf{70.55}{\scriptsize$\pm$0.42} 
            & \textbf{57.15}{\scriptsize$\pm$0.86} 
            & \textbf{65.91}{\scriptsize$\pm$0.50} 
            & 70.26{\scriptsize$\pm$1.02} 
            & \textbf{74.00}{\scriptsize$\pm$0.15} 
            & \textbf{58.26}{\scriptsize$\pm$0.74} 
            & \textbf{66.80}{\scriptsize$\pm$0.59} 
            & \textbf{66.40}{\scriptsize$\pm$0.10}  \\
			\bottomrule[1 pt]
	\end{tabular}}
	\vspace{-0.5 cm}
\end{table}

\begin{table}[t]
	\renewcommand\arraystretch{1.25}
	\newcolumntype{L}[1]{>{\raggedright\arraybackslash}p{#1}}
	\newcolumntype{C}[1]{>{\centering\arraybackslash}p{#1}}
	\newcolumntype{R}[1]{>{\raggedleft\arraybackslash}p{#1}}
	\centering
	\caption{Comparison between our proposed methods and other SOTA models on the DCASE 2024 ASD dataset.}
	\vspace{-0.1cm}
	\label{tab: SOTA}\medskip
	\resizebox{12 cm}{!}{\begin{tabular}{c c c c c c |c c c c|c}
			\toprule[1 pt]
			\multirow{2}{*}{System} & \multirow{2}{*}{Size} & \multicolumn{4}{c}{Development set} & \multicolumn{4}{c}{Evaluation set} & All \\
            {} & {} & AUC$_{\mathrm s}$ & AUC$_{\mathrm t}$ & pAUC & score & AUC$_{\mathrm s}$ & AUC$_{\mathrm t}$ & pAUC & score & score \\
            \midrule
            Official baseline\cite{harada2023first} & 267K & 65.00 & 50.28 & 52.84 & 55.35 & \textbf{71.51} & 50.58 & 51.72 & 56.50 & 55.99  \\
            
            Lv (No. 1)\cite{LvAITHU2024} & 700M & \textbf{73.97} & \textbf{72.41} & 59.16 & \textbf{67.82} & 71.03 & 73.66 & 56.70 & 66.24 & \textbf{67.02}  \\
            
            Jiang (No. 2)\cite{JiangTHUEE2024} & 360M & - & - & \textbf{59.60} & 67.67 & 69.56 & 72.34 & 56.51 & 65.36 & 66.50  \\
            
            Zhao (No. 3)\cite{ZhaoCUMT2024} & - & 60.25 & 61.50 & 53.26 & 58.10 & 68.65 & 63.91 & 54.94 & 61.96 & 59.97  \\

            AnoPatch\cite{jiang2024anopatch} & 90M & - & - & - & 62.47{\scriptsize$\pm$0.77} & - & - & - & 65.58{\scriptsize$\pm$1.12} & 63.98{\scriptsize$\pm$0.38} \\
            
            Jiang\cite{10889514} & 90M & - & - & - & 64.05{\scriptsize$\pm$0.55} & - & - & - & 66.01{\scriptsize$\pm$0.66} & 65.01{\scriptsize$\pm$0.18}  \\
            
            \midrule
            Ours & 90M & 72.25{\scriptsize$\pm$0.89} 
            & 70.55{\scriptsize$\pm$0.42} 
            & 57.15{\scriptsize$\pm$0.86} 
            & 65.91{\scriptsize$\pm$0.50} 
            & 70.26{\scriptsize$\pm$1.02} 
            & \textbf{74.00}{\scriptsize$\pm$0.15} 
            & \textbf{58.26}{\scriptsize$\pm$0.74} 
            & \textbf{66.80}{\scriptsize$\pm$0.59} 
            & 66.40{\scriptsize$\pm$0.10}  \\
			\bottomrule[1 pt]
	\end{tabular}}
	\vspace{-0.5 cm}
\end{table}

\subsubsection{Comparison with Other SOTA Models}
As shown in Table \ref{tab: SOTA}, we compare our methods and other ASD SOTA models by the performance on the DCASE 2024 ASD dataset according to the challenge rules. Here the baseline represents the official system of the DCASE 2024 ASD Challenge. Meanwhile, some of the previous SOTA systems use multiple subsystems to ensemble, so here we also list their size of overall parameters. It can be found that with our proposed methods in this paper, we achieve a new SOTA performance on the DCASE 2024 ASD evaluation dataset with fewer number parameters and a score comparable to the No. 1 and No. 2 teams on all machine types. It is noted that our methods only use one system, while No. 1 and No. 2 teams use 7 and 4 subsystems, respectively.

\section{Conclusion}
In this paper, we propose two simple yet effective methods to enhance the performance of pre-trained models in ASD. By using modified FBank audio feature, the ASD system can better capture the anomaly in machine audio, resulting in improved performance. Meanwhile, by incorporating with parameter-free and customized feature enhancement, the ASD system can better learn the crucial features of machine audio. Finally, we achieve great increases and a new SOTA performance on the DCASE 2024 ASD evaluation dataset.

%
% ---- Bibliography ----
%
% BibTeX users should specify bibliography style 'splncs04'.
% References will then be sorted and formatted in the correct style.
%
\bibliographystyle{splncs04}
\bibliography{ref.bib}
\end{document}